% =======================LETRAS HUECAS============================
\newfam\msbfam
\font\twlmsb=msbm10 at 12pt
\font\eightmsb=msbm10 at 8pt
\font\sixmsb=msbm10 at 6pt
\textfont\msbfam=\twlmsb
\scriptfont\msbfam=\eightmsb
\scriptscriptfont\msbfam=\sixmsb
\def\cj{\fam\msbfam}

\def\C{{\cj C}}

\def\R{{\cj R}}

\def\Z{{\cj Z}}
\def\I{{\cj I}}
\def\H{{\cj H}}
\centerline{\bf CHARGE CONJUGATION FROM SPACE-TIME INVERSION}

\

\centerline{\bf B. Carballo P\'erez and M. Socolovsky} 

\

\centerline{\it Instituto de Ciencias Nucleares, Universidad Nacional Aut\'onoma de M\'exico}
\centerline{\it Circuito Exterior, Ciudad Universitaria, 04510, M\'exico D. F., M\'exico} 

\

{\it We show that the $CPT$ group of the Dirac field emerges naturally from the $PT$ and $P$ (or $T$) subgroups of the Lorentz group.}

\


\

{\bf 1. $CPT$ group of the Dirac field}

\

In a recent paper (Socolovsky, 2004), it was shown that the $CPT$ group $G_{\Theta}$ ($\Theta$ for the product $\bf C *\bf P *\bf T$ of the three operators: $\bf C$, charge conjugation; $\bf P$, space inversion; $\bf T$, time reversal) of the Dirac quantum field $\hat{\psi}$, was isomorphic to the direct product of the quaternion group $Q$ and the cyclic group of two elements $\Z_2$ i.e. $$G_{\Theta}\cong Q \times \Z_2. \eqno{(1)}$$ 

The product $\bf A*\bf B$, where $\bf A$ and $\bf B$ are any of the operators $\bf C$, $\bf P$, $\bf T$, is given by $(\bf A* \bf B)\cdot \hat{\psi}=(\bf A \bf B)^\dagger \hat{\psi}(\bf A \bf B)$. $Q$ consists of the elements 1, $\iota$, $\gamma$, $\chi$ and their negatives, with the group multiplication $\iota^2=\gamma^2=\chi^2=-1$, $\iota\gamma=-\gamma\iota=\chi$, $\gamma\chi=-\chi\gamma=\iota$ and $\chi\iota=-\iota\chi=\gamma$. $\iota$, $\gamma$ and $\chi$ are the three imaginary units defining the quaternion numbers. So $G_{\Theta}$ is one of the nine non abelian groups of a total of fourteen groups with sixteen elements; only three of the non abelian groups have three generators. In particular, $G_{\Theta}\cong DC_8\times \Z_2$, where $DC_8$ is the dicyclic group of eight elements with generators $x$ and $y$; so, the generators of $DC_8\times \Z_2$ are $(x,1)$, $(y,1)$ and $(1,z)$, where $z$ is the generator of $\Z_2$ (Asche, 1989). The isomorphism with $Q$ is given by the correspondence $x\to \iota$ and $y\to \gamma$.

\

The table of $G_{\Theta}$ is the following: $$\matrix{& \bf C & \bf P & \bf T & \bf C *\bf P & \bf C *\bf T &\bf P *\bf T & \Theta \cr \bf C & 1 & \bf C *\bf P & \bf C *\bf T & \bf P & \bf T &\Theta & \bf P *\bf T \cr \bf P & \bf C *\bf P & -1 & \bf P *\bf T & -\bf C & \Theta & -\bf T & -\bf C *\bf T \cr \bf T & \bf C *\bf T & -\bf P *\bf T & -1 & -\Theta & -\bf C & \bf P & \bf C *\bf P \cr \bf C *\bf P & \bf P & -\bf C & \Theta & -1 & \bf P *\bf T & -\bf C *\bf T & -\bf T \cr \bf C *\bf T & \bf T & -\Theta & -\bf C & -\bf P *\bf T & -1 & \bf C *\bf P & \bf P \cr \bf P *\bf T & \Theta & \bf T & -\bf P & \bf C *\bf T & -\bf C *\bf P & -1 & -\bf C \cr \Theta & \bf P *\bf T & \bf C *\bf T & -\bf C *\bf P & \bf T & -\bf P & -\bf C & -1 \cr}. \eqno {(2)}$$ This table must be completed by adding to the first row and to the first column the negatives $-\bf C$, $-\bf P$,..., $-\Theta$, -1, and making the corresponding products. The isomorphism $G_\Theta \to Q \times \Z_2$ is given by 

$1\mapsto (1,1) $

$\bf C\mapsto (1,-1) $

$\bf P\mapsto (\iota,1)$

$\bf T\mapsto (\gamma,1)$

$\bf C *\bf P\mapsto (\iota,-1)$

$\bf C *\bf T\mapsto (\gamma,-1)$

$\bf P *\bf T\mapsto (\chi,1)$

$\Theta \mapsto (\chi,-1)$

$-\bf C\mapsto (-1,-1)$

$-\bf P\mapsto (-\iota,1)$

$-\bf T\mapsto (-\gamma,1)$

$-\bf C *\bf P\mapsto (-\iota,-1)$

$-\bf C *\bf T\mapsto (-\gamma,-1)$

$-\bf P *\bf T\mapsto (-\chi,1)$

$-\Theta \mapsto (-\chi,-1)$

$-1\mapsto (-1,1)$ $$\eqno{(3)}$$

where we identified $\Z_2$ with the 0-sphere $S^0=\{1,-1\}$.

\

{\bf 2. $Q$ as a subgroup of $SU(2)$}

\

With the correspondence $$\iota \mapsto \rho_1=-i\sigma_1=\pmatrix{0 & -i \cr -i & 0 \cr},$$ $$\gamma\mapsto \rho_2=-i\sigma_2=\pmatrix{0 & -1 \cr 1 & 0 \cr}, \eqno{(4)}$$ $$\chi\mapsto \rho_3=-i\sigma_3=\pmatrix{-i & 0 \cr 0 & i \cr},$$ where $\sigma_k$, $k=1,2,3$ are the Pauli matrices, $Q$ becomes isomorphic to a subgroup $H$ of $SU(2)$. (Notice that $\rho_k^\dagger=\rho_k^{-1}$ and $det \ \rho_k=+1$; also, $H$ is not an invariant subgroup of $SU(2)$.) $\Z_2$ is isomorphic to the center of $SU(2)$: $\{\I, -\I\}$ with $\I=\pmatrix{1 &0 \cr 0 & 1 \cr}$, and so $$G_\Theta \cong H\times (center \ \ of \ \ SU(2)). \eqno{(5)}$$ Since $SU(2)$ is the {\it universal covering group} of $SO(3)$: $$SU(2) \buildrel {\Phi}\over \longrightarrow SO(3) \eqno{(6)}$$ with $$\Phi\pmatrix{z & w \cr -\bar{w} & \bar{z} \cr}=\pmatrix{Re \ z^2-Re \ w^2 & Im \ z^2+Im \ w^2 & -2Re \ zw \cr Im \ z^2-Im \ w^2 & Re \ z^2+Re \ w^2 & 2Im \ zw \cr 2Re \ z\bar{w} & 2Im \ z\bar{w} & \vert z \vert^2-\vert w \vert^2 \cr},\eqno{(7)}$$ the number of finite subgroups of $SU(2)$ is the same as the number of finite subgroups of $SO(3)$, which are: the cyclic groups $C_n$, for $n=1,2,...$ ($C_2=\Z_2$); the dihedral groups $D_k$: symmetries of the rectangle for $k=2$ and of the regular polygons for $k\geq 3$; and the rotational symmetry groups: $T$ of the tetrahedron, $O$ of the cube and the octahedron, and $I$ of the dodecahedron and the icosahedron (Sternberg, 1994). If $A$ is a finite subgroup of $SO(3)$, then $\tilde{A}=\Phi^{-1}(A)$ has twice the number of elements of $A$. 

\

{\bf 3. $\Phi(H)$ in $SO(3)$}

\

$\Phi(H)$ has four elements and so the unique candidates for it are groups isomorphic to $C_4$ and $D_2\cong \Z_2\times\Z_2$, the Klein group. A simple application of (7) to the elements of $H$ leads to $$\Phi (H)=\{I, \ R_x(\pi), \ R_y(\pi), \ R_z(\pi)\} \eqno{(8)}$$ with $$R_x(\pi)=\pmatrix{1 & 0 & 0 \cr 0 & -1 & 0 \cr 0 & 0 & -1 \cr}, \ R_y(\pi)=\pmatrix{-1 & 0 &0 \cr 0 & 1 & 0 \cr 0 & 0 & -1 \cr}, \ and \ R_z(\pi)=\pmatrix{-1 & 0 & 0 \cr 0 & -1 & 0 \cr 0 & 0 & 1 \cr}, \eqno{(9)}$$ the rotations in $\pi$ around the axis $x$, $y$ and $z$ respectively, and $I$=$\pmatrix{1 & 0 & 0 \cr 0 & 1 & 0 \cr 0 & 0 & 1 \cr}$ the unit matrix in $SO(3)$. It is immediately verified that the multiplication table of $\Phi(H)\subset SO(3)$ is the same as for $D_2$, namely $$\matrix{& a & b & c \cr a & e & c & b \cr b & c & e & a \cr c & b & a & e \cr} \eqno{(10)}$$ with the correspondence $a\mapsto R_x(\pi)$, $b\mapsto R_y(\pi)$ and $c\mapsto R_z(\pi)$. Then we have $$ G_\Theta \cong \Phi^{-1}(D_2)\times \Z_2. \eqno{(11)}$$

\

{\bf 4. Parity and time reversal}

\ 

Within the Lorentz group $O(3,1)$, the transformations of {\it parity} and {\it time reversal} are given by the matrices $$P=\pmatrix{1 & 0 & 0 & 0 \cr 0 & -1 & 0 & 0 \cr 0 & 0 & -1 & 0 \cr 0 & 0 & 0 & -1 \cr} \ \ and \ \ T=\pmatrix{-1 & 0 & 0 & 0 \cr 0 & 1 & 0 & 0 \cr 0 & 0 & 1 & 0 \cr 0 & 0 & 0 & 1 \cr} \eqno{(12)}$$ respectively. Together with the 4$\times$4 unit matrix $E$ and their product $PT$, they lead to the subgroup of the Lorentz group with multiplication table $$\matrix{& P & T & PT \cr P & E & PT & T \cr T & PT & E & P \cr PT & T & P & E \cr}. \eqno{(13)}$$ We call this group {\it the PT- group} of the Lorentz group. From (13) and (10), this group is isomorphic to $D_2$. On the other hand, $P$ or $T$ separately, together with the unit 4$\times$4 matrix $E$, give rise to the group $\Z_2$. Then we obtain the desired result: $$G_\Theta\cong \Phi^{-1}(<\{P,T\}>)\times <\{P\}> \eqno{(14a)}$$ or $$G_\Theta\cong \Phi^{-1}(<\{P,T\}>)\times <\{T\}>. \eqno{(14b)}$$ I.e., $G_\Theta$, which in the present context is a group acting at the quantum field level that includes the charge conjugation operator for the electron field $\hat{\psi}$, emerges in a natural way from the pure space-time (and therefore classical) $PT$-group and its $P$- or $T$- subgroups.

\

{\bf 5. Discussion}

\

The above result suggests that the Minkowskian space-time structure of special relativity, in particular the unconnected component of its symmetry group, the real Lorentz group $O(3,1)$, implies the existence of the $CPT$ group as a whole, and therefore the existence of the charge conjugation transformation, and thus the proper existence of antiparticles. The above conclusion (at least at the level of the electron field) is supported by the following consideration: even though the charge conjugation operation does not belong to the Lorentz group, at the level of the Dirac equation, the matrix $$C=\pm i\gamma_2\gamma_0=\pmatrix{0 & 0 & 0 & \mp 1 \cr 0 & 0 & \pm1 & 0 \cr 0 & \mp 1 & 0 & 0 \cr \pm 1 & 0 & 0 & 0 \cr} \eqno{(15)}$$ is an element of the Dirac algebra ($\cong \C(4)$), which is the complexification of any of the two non isomorphic real Clifford algebras of Minkowski space-time: $\H(2)$ (2$\times$2 quaternion matrices for the metric $diag \ (1,-1,-1,-1)$) and $\R (4)$ (4$\times$4 real matrices for the metric $diag \ (-1,1,1,1)$) (Lawson and Michelsohn, 1989).

\

Also, there is a sort of analogy between our result and the proof of the $CPT$ theorem by Jost (Jost, 1957), who based his demonstration on the existence of the $PT$ transformation as an element of the connected component of the complexification of the Lorentz group, and arguments of analytic continuation in field theory. See also Greenberg (Greenberg, 2006).

\

{\bf Acknowledgement}

\

This work was partially supported by the project PAPIIT IN 113607, DGAPA-UNAM, M\'exico.

\

{\bf References}

\

Asche, D. (1989). {\it An Introduction to Groups}, Adam Hilger, Bristol.

\

Greenberg, O. W. (2006). Why is $CPT$ fundamental?, {\it Foundations of Physics} {\bf 36}, 1535-1553; arXiv: hep-ph/0309309 v1.

\

Jost, R. (1957). Eine Bemerkung zum $CPT$-Theorem, {\it Helvetical Physical Acta} {\bf 30}, 409-416. 

\

Lawson, H. B. Jr. and M. L. Michelsohn. (1989). {\it Spin Geometry}, Princeton University Press, Princeton.

\

Socolovsky, M. (2004). The CPT Group of the Dirac Field, {\it International Journal of Theoretical Physics} {\bf 43}, 1941-1967; arXiv: math-ph/0404038 v2.

\

Sternberg, S. (1994). {\it Group Theory and Physics}, Cambridge University Press, Cambridge.

\

\

\

e-mails: brendacp@nucleares.unam.mx ; socolovs@nucleares.unam.mx

\end

\
As is well known, the magnetic Aharonov-Bohm (A-B) effect $^{1,2}$ is a gauge invariant, non local quantum phenomenon, with gauge group $U(1)$, which takes place in a non simply connected space. It involves a magnetic field in a region where an electrically charged particle obeying the Schroedinger equation cannot enter, i.e. the ordinary 3-dimensional space minus the space occupied by the solenoid producing the field; in the ideal mathematical limit, the solenoid is replaced by a flux line. Locally, the particle couples to the magnetic potential $\vec{A}$ but not to the magnetic field $\vec{B}$; however, the effect is gauge invariant since it only depends on the flux of $\vec{B}$ inside the solenoid. 

\

The fibre bundle theoretic description of this kind of phenomena has proved to be very useful to obtain a more profound insight into the relation between physical processes and pure mathematics. $^3$ In the present case, since by symmetry, the dimension along the flux line can be ignored, the problem reduces to the effect on the charged particle of an abelian connection in a $U(1)$-bundle with base space the plane minus a point. The wave function representing the particle is a section of an associated vector bundle. As shown in ref. 4, the bundle turns out to be trivial and then its total space is isomorphic to the product $(\R^2-\{point\})\times U(1)$. Since $\R^2$ is topologically equivalent to an open disk, and $U(1)$ is the unit circle, the bundle structure is summarized by $$U(1)\to {T}^2_{\circ*}\to {D}^2_{\circ*} \eqno{(1)}$$ where $D^2_{\circ*}$ is the open disk minus a point and $T^2_{\circ*}$ is the open solid 2-torus minus a circle. 

\

As suggested by Wu and Yang $^6$, Yang-Mills fields can give rise to non abelian A-B effects. In ref. 6 the authors studied an $SU(2)$ gauge configuration leading to an A-B effect; later, several authors studied the effect with gauge groups $SU(3)$ $^7$ and $U(N)$ $^8$. Also, in refs. 9-14 the effect was studied in the context of gravitation theory. In all these examples, as in the  magnetic case, there is a principal bundle structure $$\xi: G\to P\buildrel {\pi_G}\over\longrightarrow M \eqno{(2)}$$ whose total space $P$, where the connection giving rise to the effect lies, is however never specified. We shall restrict ourselves to connections $\omega$ giving rise to $n$ ($n=1,2,3,...$) flux lines in $\R^3$. In this note we prove the following: 

\

{\it Theorem}: Let $G$ be a path connected topological group (for example a connected Lie group) and $\xi$ a continuous principal $G$-bundle over $\R^3 - \{n \ parallel \ lines\}$, $n=1,2,3,...$ Then the bundle $\xi$ is trivial, i.e. isomorphic to the product bundle.

\

{\it Proof of the theorem}

\

The classification of bundles over $\R^3 \backslash \{n$ parallel lines$\}$ is the same as that over $\R^2 \backslash \{n$ points$\}$ which is topologically equivalent to $D^2_\circ \backslash \{n$ points $\}\equiv D^2_{\circ * n}$. Denote this set of points by $\{b_1,...,b_n\}$. By symmetry along the dimension of the flux tubes, $D^2_{\circ*n}$ is the space where it can be considered that the charged particles move. 

Let $x_0$ be a point in $D^2_{\circ*n}$. We construct a bouquet of $n$ loops $\gamma_1$, $\gamma_2$,..., $\gamma_n$ through $x_0$, with the k-th loop sorrounding the point $b_k$, $k=1,2,...,n$. This space is homeomorphic to the wedge product (or reduced join) $S^1\vee...\vee S^1\equiv \vee_n S^1\equiv S^1_{(1)}\vee...\vee S^1_{(n)}$ of $n$ circles $^{15}$, and the classification of bundles over $D^2_{\circ*n}$ is the same as that over $\vee_n S^1$, namely $${\cal B}_{D^2_{\circ*n}}(G)={\cal B}_{\vee_n S^1}(G) \eqno{(3)}$$ where ${\cal B}_M(G)$ is the set of isomorphism classes of $G$-bundles over $M$ $^{16}$. 

By explicit construction we shall prove that, up to isomorphism, the unique $G$-bundle over $\vee_n S^1$ is the product bundle $\vee_n S^1\times G$ (which is a purely topological result). With this aim, we cover the circle $S_{(k)}^1$ with two open sets $U_{k+}$ and $U_{k-}$ such that $$U_{k+}\cap U_{k-}\simeq\{x_0,a_k\}, \ k=1,...,n, \eqno{(4)}$$ $$U_{i+}\cap U_{j+}\simeq U_{i-}\cap U_{j-}\simeq U_{i+}\cap U_{j-}\simeq \{x_0\}, \ i,j=1,...,n, \ i\neq j \eqno{(5)}$$ where $\simeq$ denotes homotopy equivalence, and $a_k\in S^1_{(k)}$ with $a_k \neq x_0$. We then have $$2\times\pmatrix{2n \cr 2 \cr}+2n={{2\times(2n)!}\over {(2n-2)!2!}}+2n=4n^2 \eqno{(6)}$$ transition functions $$g_{\alpha, \beta}:U_\alpha \cap U_\beta \to G \eqno{(7)}$$ with $g_{\beta,\alpha}=g_{\alpha,\beta}^{-1}$. Up to homotopy they are given by $$g_{k+,k-}:\{x_0,a_k\}\to G, \ x_0 \mapsto g_{0k}, \ a_k\mapsto g_k, \ k=1,...,n \eqno{(8)}$$ and $$g_{i+,j+},g_{i-,j-},g_{i+,j-}:\{x_0\}\to G, \ i,j=1,...,n, \ i\neq j,$$ $$x_0\mapsto g_{ij++}, \ x_0\mapsto g_{ij--}, \ x_0\mapsto g_{ij+-}. \eqno{(9)}$$ The $2n$ transition functions $g_{i+,i+}$, $g_{i-,i-}$, $i\in\{1,...,n\}$, give the identity in $G$. The number of cocycle relations $g_{\beta,\alpha}g_{\alpha,\gamma}=g_{\beta,\gamma}$ on the $2n(2n-1)$ non trivial transition functions $g_{\mu,\nu}$ is $\pmatrix{2n \cr 3\cr}$=${{n(2n-1)(2n-2)}\over{3}}$.

Let $g^\prime_{k+,k-}:\{x_0,a_k\}\to G$ be given by $g^\prime_{k+,k-}(x_0)=g^\prime_{0k}$, $g^\prime_{k+,k-}(a_k)=g^\prime_k$. Since $G$ is path connected, there exist continuous paths $$c^{(k)}_0:[0,1]\to G, \ t\mapsto c_0^{(k)}(t) \ \ with \ \ c_0^{(k)}(0)=g_{0k}, \ c_0^{(k)}(1)=g^\prime_{0k}$$ and $$c_k^{(k)}:[0,1]\to G, \ t\mapsto c_k^{(k)}(t) \ \ with \ \ c_k^{(k)}(0)=g_k, \ c_k^{(k)}(1)=g^\prime_k.$$ Then the continuous function $$H_k:\{x_0,a_k\}\times [0,1]\to G$$ given by $$H_k(x_0,t)=c_0^{(k)}(t), \ \ H_k(a_k,t)=c_k^{(k)}(t) \eqno{(10)}$$ is a homotopy between $g_{k+,k-}$ and $g^\prime_{k+,k-}$ since $$H_k(x_0,0)=g_{0k} \ \ and \ \  H_k(a_k,0)=g_k$$ i.e. $$H_k\vert_{\{x_0,a_k\}\times \{0\}}=g_{k+,k-}$$ and $$H_k(x_0,1)=g^\prime_{0k} \ \ and \ \ H_k(a_k,1)=g^\prime_k$$ i.e. $$H_k\vert_{\{x_0,a_k\}\times\{1\}}=g^\prime_{k+,k-}.$$ Then the homotopy class of maps from $\{x_0,a_k\}$ to $G$ has only one element, namely $[g_{k+,k-}]_{\sim}$ i.e. $$[\{x_0,a_k\},G]_{\sim}=\{[g_{k+,k-}]_{\sim}\}. \eqno{(11)}$$

Similarly, let $g_{i\mu,j\nu}$ with $\mu,\nu \in\{+,-\}$ be any of the functions in (9), and let $g^\prime_{i\mu,j\nu}:\{x_0\}\to G$ be given by $$g^\prime_{i\mu,j\nu}(x_0)=g^\prime_{ij,\mu\nu}; \eqno{(12)}$$ then the map $$H_{i\mu,j\nu}:\{x_0\}\times[0,1]\to G, \ \ H_{i\mu,j\nu}(x_0,t)=c_{i\mu,j\nu}(t) \eqno{(13)}$$ with $c_{i\mu,j\nu}:[0,1]\to G$ a continuous path in $G$ satisfying $c_{i\mu,j\nu}(0)=g_{ij\mu\nu}$ and $c_{i\mu,j\nu}(1)=g^\prime_{ij\mu\nu}$, is a homotopy between $g_{i\mu,j\nu}$ and $g^\prime_{i\mu,j\nu}$ i.e. $$[\{x_0\},G]_{\sim}=\{[g_{i\mu,j\nu}]_{\sim} \}. \eqno{(14)}$$ It is then easy to show that, if we define (constant) functions $$\Lambda_{k\mu}:U_{k\mu}\to G, \ \ p \to \lambda_{k\mu}, \ \ k=1,...,n, \ \ \mu=+,-,\eqno{(15)}$$ then $$g^\prime_{k+,k-}(x_0)=\lambda_{k-}g_{k+,k-}(x_0)\lambda_{k+}^{-1}, \eqno{(16)}$$ $$g^\prime_{k+,k-}(a_k)=\lambda_{k-}g_{k+,k-}(a_k)\lambda_{k+}^{-1}, \eqno{(17)}$$ for $k=1,...,n$, and $$g^\prime_{i\mu,j\nu}(x_0)=\lambda_{j\nu}g_{i\mu,j\nu}(x_0)\lambda_{i\mu}^{-1} \eqno{(18)}$$ for $i,j\in\{1,...,n\}$, $i\neq j$, $\mu,\nu\in\{+,-\}$. In fact, for arbitrary two sets of homotopic transition functions $g_{\beta,\alpha}^\prime \sim g_{\beta,\alpha}$, one has continuous maps $H:(U_\beta \cap U_\alpha)\times [0,1]\to G$ such that $H(x,0)=g_{\beta,\alpha}(x)$ and $H(x,1)=g_{\beta,\alpha}^\prime(x)$. Given a map $\Lambda_\alpha:U_\beta \cap U_\alpha \to G$, $x\to \Lambda(x)=\lambda_\alpha$, one defines $\bar{H}(x,1)=H(x,1)\Lambda_\alpha(x)$ and therefore $\bar{H}(x,1)=g_{\beta,\alpha}^\prime(x)\Lambda_{\alpha}(x)$. Now, one defines $\Lambda_\beta:U_\beta \cap U_\alpha \to G$ through $\Lambda_{\beta}(x)=\bar{H}(x,1)H(x,0)^{-1}=g_{\beta,\alpha}^\prime(x)\Lambda_{\alpha}(x)g_{\beta,\alpha}(x)^{-1}$ i.e. $g_{\beta,\alpha}^\prime(x)=\Lambda_\beta(x)g_{\beta,\alpha}(x)\Lambda_\alpha(x)^{-1}$ (*) for all $x\in U_\beta\cap U_\alpha$. In our case, for $U_{k+}\cap U_{k-}\simeq \{x_0,a_k\}$, the formulae corresponding to (*) are $$g_{k+,k-}^\prime(x_0)=\Lambda_{k-}(x_0)g_{k+,k-}(x_0)\Lambda_{k+}(x_0)^{-1}$$ and $$g_{k+,k-}^\prime(a_k)=\Lambda_{k-}(a_k)g_{k+,k-}(a_k)\Lambda_{k+}(a_k)^{-1}.$$ Extending continuously (as constants) $\Lambda_{k+}$ and $\Lambda_{k-}$ respectively to the open sets $U_{k+}$ and $U_{k-}$: $$\Lambda_{k+}:U_{k+}\to G, \ p\mapsto \lambda_{k+}, \ \ \Lambda_{k-}:U_{k-}\to G, \ q\mapsto \lambda_{k-}$$ (in particular one has $\Lambda_{k+}(x_0)=\Lambda_{k+}(a_k)=\lambda_{k+}$ and $\Lambda_{k-}(x_0)=\Lambda_{k-}(a_k)=\lambda_{k-}$), one obtains equations (16) and (17). Proceeding similarly for the cases $U_{i\mu}\cap U_{j\nu}\simeq\{x_0\}$, one gets equation (18).

Then, up to isomorphisms, and according to a general theorem for coordinate bundles $^{17}$, there is a unique $G$-bundle generated by the transition functions given by the equations (8) and (9), namely the product bundle. We then have $${\cal B}_{\vee_n S^1}(G)=\{[G\to\vee_n S^1 \times G\to \vee_n S^1]\}, \eqno{(19)}$$ where $[ \ \ \ ]$ denotes here the equivalence class of bundles isomorphic to the product bundle.    QED 

\

For the cases of the examples in refs. 1, 6 and 7, we have, respectively, $${\cal B}_{D^2_{\circ *}}(U(1))=\{[U(1)\to D^2_{\circ *}\times U(1)\to D^2_{\circ *}]\}, \eqno{(20)}$$ $${\cal B}_{D^2_{\circ *}}(SU(2))=\{[SU(2)\to D^2_{\circ *}\times SU(2)\to D^2_{\circ *}]\}, \eqno{(21)}$$ and $${\cal B}_{D^2_{\circ *2}}(SU(3))=\{[SU(3)\to D^2_{\circ *2}\times SU(3)\to D^2_{\circ *2}]\}. \eqno{(22)}$$ 

\

In the gravitational case, the gauge group is $SL(2,\C)$ $^{18}$, which is the universal covering group of the connected component of the Lorentz group $L^\uparrow_+$; then, for weak gravitational fields with a distribution of $n$ gravitomagnetic flux lines as above we would have the trivial bundle $$SL(2,\C)\to D^2_{\circ *n}\times SL(2,\C)\to D^2_{\circ *n}. \eqno{(23)}$$

\

We want to stress that, even if the A-B connection is flat (though not exact), one of the sufficient conditions for the automatic triviality of the A-B bundle fails: though paracompact, the bouquet $\vee_n S^1$ is not simply connected. (See corollary 9.2. in ref. 21, p. 92.) In addition, there is not a priori any physical reason why, under the specified conditions on the ``magnetic flux'', the corresponding A-B  bundle should be trivial.

\

It is interesting to notice that there are two other fibre bundles related to the A-B effect. The first bundle is the {\it universal covering space} $^{19}$ of the base manifold (``laboratory'' or physical space where the particles coupled to the A-B potential move) which is the $\pi_1(M;x_0)$-(non trivial) bundle $\xi_c:\tilde{M}\buildrel {\pi}\over\longrightarrow M$, where $\pi_1(M;x_0)$ ($\equiv \pi_1(M)$ if $M$ is connected) is the fundamental group of $M$. In our case, $$\pi_1(\vee_n S^1;x_0)\cong \pi_1(\R^2\backslash \{n \ points\};x_0)\cong <\{c_1,...,c_n\}>$$ is the freely generated group with $n$ non commuting generators $c_1,...c_n$. In particular, for the original abelian A-B effect with group $U(1)$, $\tilde{\R^{2*}}=RS(Log)$: the Riemann surface of the logarithm, and $\pi_1(\R^{2*})\cong \Z$. 

\

The {\it particle propagator} in $M$, $K(x^{\prime\prime},t^{\prime\prime}; x^\prime,t^\prime)$ with $t^{\prime\prime}>t^\prime$, is a sum of homotopy propagators $^7$ multiplied by corresponding gauge factors $^{20}$: the former are given by unrestricted path integrals computed in $\tilde{M}$, the paths in these path integrals project onto the corresponding homotopy classes of paths in the non simply connected space $M$; the latter are Wilson loops given by $$Texp\int_{\pi(c)}\vec{A}\cdot d\vec{l}$$ where $T$ denotes time order, $\vec{A}$ is the A-B potential, and $c$ is a loop in $\tilde{M}$ beginning  and ending respectively at $y_0$ and $y^{\prime\prime}$ in $\pi^{-1}(\{x^{\prime\prime}\})\buildrel{\Psi}\over\cong \pi_1(M)$, with $y_0$ fixed and arbitrary. Then one has the group homomorphism (many-to-one or one-to-one) $$y^{\prime\prime}\buildrel {\Psi}\over \longrightarrow \Psi(y^{\prime\prime})\buildrel{\varphi}\over\longrightarrow Texp\int_{\pi(c)}\vec{A}\cdot d\vec{l} \ \ \in G \eqno{(24)}$$ whose image in $G$, $\varphi(\pi_1(M))$, responsible for the A-B effect, is the {\it holonomy} of the connection. $^{21}$

\

The second bundle is the {\it associated complex vector bundle} $\xi_{\C ^m}: \C ^m - P_{\C ^m}\buildrel {\pi_{\C ^m}}\over \longrightarrow M$ ($m=2s+1$ is the dimension of the spinor space and $s$ is the spin; for scalar particles $m=1$), where $$P_{\C^m}=(M\times G)\times_G \C^m=\{[((x,g),\vec{z})]\}_{((x,g),\vec{z})\in(M\times G)\times_G\C^m},$$ $$[((x,g),\vec{z})]=\{((x,gg^\prime),g^{\prime -1}\vec{z})\}_{g^\prime \in G}. \eqno{(25)}$$ $\xi_{\C^m}$ is trivial since $\xi$ is trivial, and the quantum mechanical {\it wave functions} of the particles are global sections of $\xi_{\C^m}$: $$\psi\in \Gamma(\xi_{\C^m})$$ i.e. $\psi:M\to P_{\C^m}$ with $\pi_{\C^m}\circ \psi= Id_M$. 

\

Notice that while the propagator is computed in $\xi_c$, the wave function lies in $\xi_{\C^m}$, with $$\psi(x^{\prime\prime},t^{\prime\prime})=\int_M dx^\prime K(x^{\prime\prime},t^{\prime\prime};x^\prime,t^\prime)\psi(x^\prime,t^\prime).\eqno{(26)}$$ If $\omega$ is the A-B connection in $P$, then the coupling $\omega-\psi$ is the covariant derivative $$\nabla_V^\omega \psi=\psi_{V^\uparrow(\gamma_\psi)}\in \Gamma(\xi_{\C^m}), \eqno{(27)}$$ where $V$ is a vector field in $M$, $V^\uparrow$ its horizontal lifting in $P$ by $\omega$, $\gamma_\psi$ and $V^\uparrow(\gamma_\psi)$ are equivariant functions from $P$ to $\C^m$ with $\gamma_\psi(p)=\vec{z}$  where $\psi(\pi_G(p))=[p,\vec{z}]$, and $\psi_{V^\uparrow(\gamma_\psi)}(x)=[p,V^\uparrow(\gamma_\psi)(p)]$ for any $p\in \pi_G^{-1}(\{x\})$. Locally, of course, $\nabla_V^\omega\psi$ reproduces the usual minimal coupling between $\vec{A}$ and $\psi$.

\

Finally, though $\varphi:\pi_1(M)\to G$ is a group homomorphism, and $\tilde{M}\buildrel\ {f}\over \longrightarrow M \times G$ given by $f(y)=(\pi(y),1)$ is a canonical map, there is no bundle map between $\xi_c$ and $\xi$: the pair of functions $(f\times\varphi,f)$ is {\it not} a principal bundle homomorphism.   

\

In summary, the three bundles are related by the following diagram:

$$\matrix{\pi_1(M) & \buildrel{\varphi}\over\longrightarrow & G & & & &  \C^m \cr
          \downarrow & &  \downarrow & & & &  \vert \cr
          \tilde{M} & \buildrel{f}\over \longrightarrow & M\times G & &  \buildrel{\iota}\over \longrightarrow & &  (M\times G)\times_G\C^m \cr
          \downarrow \pi & & \downarrow \pi_G & & & & \downarrow \pi_{\C^m}\uparrow \psi \uparrow \nabla^\omega _V \psi \cr
          M & = & M & & = & & M \cr}$$ where $\iota$ is the canonical injection of the bundle $\xi$ into its associated bundle i.e. $\iota(p)=[p,0]$. 

\

{\bf Acknowledgments}

\

This work was partially supported by the grant PAPIIT-UNAM IN103505. M. S. thanks for hospitality to the Instituto de Astronom\'\i a y F\'\i sica del Espacio (UBA-CONICET, Argentina), and the University of Valencia, Spain, where part of this work was done. 

\

{\bf References}

\

1. Aharonov, Y. and Bohm, D.(1959). Significance of Electromagnetic Potentials in Quantum Theory, {\it Physical Review}, {\bf 115}, 485-491.

\

2. Chambers, R. G. (1960). Shift of an Electron Interference Pattern by Enclosed Magnetic Flux, {\it Physical Review Letters}, {\bf 5}, 3-5.

\

3. Daniel, M. and Viallet, C. M. (1980). The Geometrical Setting of Gauge Theories of the Yang-Mills Type, {\it Reviews of Modern Physics}, {\bf 52}, pp. 175-197.

\

4. Aguilar, M. A. and Socolovsky, M. (2002). Aharonov-Bohm Effect, Flat Connections, and Green's Theorem, {\it International Journal of Theoretical Physics}, {\bf 41}, 839-860. (For a review see, e.g. ref. 5.)

\

5. Socolovsky, M. (2006). Aharonov-Bohm Effect, in {\it Encyclopedia of Mathematical Physics}, Elsevier, Amsterdam, pp. 191-198.

\

6. Wu, T. T. and Yang, C. N. (1975). Concept of non Integrable Phase Factors and Global Formulation of Gauge Fields, {\it Physical Review D}, {\bf 12}, 3845-3857.

\

7. Sundrum, R. and Tassie, L. J. (1986). Non-Abelian Aharonov-Bohm Effects, Feynman Paths, and Topology, {\it Journal of Mathematical Physics}, {\bf 27}, 1566-1570.

\

8. Botelho, L. C. L. and de Mello, J. C. (1987). A Non-Abelian Aharonov-Bohm Effect in the Framework of Feynman Pseudoclassical Path Integrals, {\it Journal of Physics A: Math. Gen.}, {\bf 20}, 2217-2219. 

\

9. Harris, E. G. (1996). The gravitational Aharonov-Bohm effect with photons, {\it American Journal of Physics}, {\bf 64}, 378-383.

\

10. Zeilinger, A., Horne, M. A. and Shull, C. G. (1983). Search for unorthodox phenomena by neutron interference experiments, {\it Proceedings International Symposium of Foundations of Quantum Mechanics}, Tokio, pp. 289-293.

\

11. Ho, Vu B. and Morgan, M. J. (1994). An experiment to test the gravitational Aharonov-Bohm effect, {\it Australian Journal of Physics}, {\bf 47}, 245-252.

\

12. Bezerra, V. B. (1987). Gravitational analogue of the Aharonov-Bohm effect in four and three dimensions, {\it Physical Review D}, {\bf 35}, 2031-2033.

\

13. Wisnivesky, D. and Aharonov, Y. (1967). Nonlocal effects in classical and quantum theories, {\it Annals of Physics}, {\bf 45}, 479-492.

\

14. Corichi, A. and Pierri, M. (1995). Gravity and Geometric Phases, {\it Physical Review D}, {\bf 51}, 5870-5875.

\

15. Greenberg, M. J. and Harper, J. R. (1981). {\it Algebraic Topology. A First Course}, Addison-Wesley, Redwood City, p. 126.  

\

16. Nash, C. and Sen, S. (1983). {\it Topology and Geometry for Physicists}, Academic Press, London, p. 262.

\

17. Idem 16., pp. 147-148.

\

18. Naber, G. L. (2000). {\it Topology, Geometry, and Gauge Fields, Interactions}, Springer-Verlag, New York, pp. 193-197.

\

19. Massey, W. S. (1991). {\it A Basic Course in Algebraic Topology}, GTM 56, Springer, New York, p. 132. 

\

20. Schulman, L. S. (1981). {\it Techniques and Applications of Path Integration}, Wiley, New York, pp. 205-207.  

\

21. Kobayashi, S and Nomizu, K. (1963). {\it Foundations of Differential Geometry, Vol. 1}, Wiley, New York, p. 71.

\

\

\

\

\

\

\

e-mails: 

\

rshuerfanob@unal.edu.co

\

alicia@nucleares.unam.mx

\

socolovs@nucleares.unam.mx

\end

\

2. {\bf Galilean group, its universal covering group, and spinors}

\

The connected component of the galilean group $G_0$ consists of the set of 4$\times$4 matrices $$g=\pmatrix{R&\vec{V}\cr 0&1\cr} \eqno{(6)}$$ with $R$ in the 3-dimensional rotation group $SO(3)$, boost velocity $\vec{V}$ in $\R^3$, composition law $$g_2g_1=\pmatrix{R_2 &\vec{V}_2 \cr 0 & 1 \cr}\pmatrix{R_1 & \vec{V}_1 \cr 0 &1\cr}=\pmatrix{R_2R_1 & \vec{V}_2+R_2\vec{V}_1 \cr 0 & 1 \cr}, \eqno{(6a)}$$ identity $$\pmatrix{I & 0 \cr 0 &1 \cr}, \ \ \ I=\pmatrix{1&0&0\cr 0&1&0 \cr 0&0&1}, \eqno{(6b)}$$ and inverse $$\pmatrix{R & \vec{V}\cr 0 &1}^{-1}=\pmatrix{R^{-1} & -R^{-1}\vec{V} \cr 0 & 1}. \eqno{(6c)}$$ $G_0$ is a non abelian, non compact, connected but non simply connected six dimensional Lie group; like the connected component of the Lorentz group, its topology is that of the cartesian product of the real projective space with ordinary 3-space {\it i.e.} of $\R P^3\times \R ^3$. The action of $G_0$ on spacetime is given by $$G_0\times \R ^4\to \R ^4, \ (g,\pmatrix{\vec{x}^\prime \cr t^\prime \cr})\mapsto \pmatrix{\vec{x}\cr t \cr}=g\pmatrix{\vec{x}^\prime \cr t^\prime \cr}=\pmatrix{R\vec{x}^\prime+\vec{V}t^\prime \cr t^\prime}. \eqno{(7)}$$ Since one has the action $$\mu:SO(3)\times \R ^3 \to \R ^3, \ (R,\vec{x})\mapsto R\vec{x}, \eqno{(8)}$$then $G_0$ is isomorphic to the semidirect sum $\R ^3 \times_{\mu}SO(3)$: $\pmatrix{R & \vec{V} \cr 0 & 1 \cr}\mapsto (\vec{V},R)$ with composition law $$(\vec{V}^\prime,R^\prime)(\vec{V},R)=(\vec{V}^\prime+R^\prime\vec{V},R^\prime R). \eqno{(8a)}$$

\ 

The {\it universal covering group} of $G_0$ is given by the $\Z_2$-bundle $$\Z_2\to \hat{G}_0 \buildrel {\Pi}\over \longrightarrow G_0 \eqno{(9)}$$ where $$\hat{G}_0=\{\hat{g}=\pmatrix{T & \vec{V} \cr 0 & 1 \cr}, \ T\in SU(2), \ \vec{V}\in \R ^3\}, \eqno{(9a)}$$ and $\Pi$ is the $2\to 1$ group homomorphism $$\Pi(\hat{g})=\pmatrix{\pi(T) & \vec{V} \cr 0 & 1 \cr} \eqno{(9b)}$$ with $\pi:SU(2) \to SO(3)$ the well known projection $$\pi\pmatrix{z & w \cr -\bar{w} & \bar{z} \cr}=\pmatrix{Rez^2-Rew^2 & Imz^2+Imw^2 & -2Rezw \cr -Imz^2+Imw^2 & Rez^2+Rew^2 & 2Imzw \cr 2Rez\bar{w} & 2 Imz\bar{w} & \vert z\vert ^2 -\vert w \vert ^2 \cr}. \eqno{(9c)}$$ $\hat{G}_0$ is simply connected and has the topology of $S^3\times \R ^3$. Since $SU(2)$ acts on $\R ^3$: $$\hat{\mu}:SU(2)\times \R ^3\to \R ^3, \ (T,\vec{V})\mapsto \pi(T)\vec{V},\eqno{(10)}$$ one has the group isomorphism $$\hat{G}_0\ni\pmatrix{T & \vec{V} \cr 0 & 1 \cr}\mapsto (\vec{V},T) \in \R ^3 \times _{\hat{\mu}}SU(2); \eqno{(11)}$$ the composition law in $\hat{G}_0$ is given by $$\pmatrix{T^\prime & \vec{V}^\prime \cr 0 & 1 \cr}\pmatrix{T & \vec{V} \cr 0 & 1 \cr}= \pmatrix{T^\prime T & \vec{V}^\prime+\pi(T^\prime)\vec{V} \cr 0 & 1 \cr}, \eqno{(12)}$$ while the identity and inverse are respectively given by $$\pmatrix{I & 0 \cr 0 & 1 \cr}, \ I=\pmatrix{1 & 0 \cr 0 & 1 \cr} \eqno{(12a)}$$ and $$\pmatrix{T & \vec{V} \cr 0 & 1 \cr}^{-1}=\pmatrix{T^{-1} & -\pi(T^{-1})\vec{V} \cr 0 & 1 \cr}. \eqno{(12b)}
$$ 

\

Turning back to physics, for each mass value $m>0$, $\hat{G}_0$ acts on the infinite dimensional Hilbert space ${\cal L}^2_1$ of continuously differentiable and square integrable $\C ^2$-valued functions $\pmatrix{u \cr v \cr}$ on $\R ^4$, the Schroedinger-Pauli spinors. This action is defined as follows: $^5$ $$\hat{\mu}_m:\hat{G}_0\times {\cal L}^2_1 \to {\cal L}^2_1, \ (\pmatrix{T & \vec{V} \cr 0 & 1 \cr},\pmatrix{u \cr v \cr})\mapsto \pmatrix{T & \vec{V} \cr 1 & 0 \cr}\cdot \pmatrix{u & \cr v \cr}: \R ^4 \to \C ^2,$$ $$\pmatrix{\vec{x} \cr t \cr}\mapsto \pmatrix{T & \vec{V} \cr 0 & 1 \cr}\cdot \pmatrix{u \cr v \cr}\pmatrix{\vec{x} \cr t \cr}= e^{{{-im}\over {\hbar}}(\vec{V}\cdot\vec{x}+{{1}\over{2}}\vert \vec{V}\vert ^2 t)}T\pmatrix{u(\pi(T)\vec{x}+\vec{V}t,t) \cr v(\pi(T)\vec{x}+\vec{V}t,t)}. \eqno{(13)}$$ $\hat{\mu}_m$ is equivalent to the {\it representation} $$\tilde{\hat{\mu}}_m:\hat{G}_0 \to End({\cal L}^2_1), \ \tilde{\hat{\mu}}_m(\hat{g})(\pmatrix{u \cr v & \cr})=\hat{g}\cdot \pmatrix{u \cr v \cr}. \eqno{(13.a)}$$ At each $t$ one has the inner product $$(\pmatrix{u_2 \cr v_2 \cr},\pmatrix{u_1 \cr v_1 \cr})(t)=\int d^3\vec{x}(\bar{u}_2(\vec{x},t)u_1(\vec{x},t)+\bar{v}_2(\vec{x},t)v_1(\vec{x},t)) \eqno{(14a)}$$ and the norm $$\vert \vert \pmatrix{u \cr v \cr} \vert \vert ^2 (t)=(\pmatrix{u \cr v \cr}, \pmatrix{u \cr v \cr})(t)=\int d^3\vec{x}(\vert u(\vec{x},t)\vert ^2 + \vert v(\vec{x},t)\vert ^2). \eqno{(14b)}$$ The galilean transformation of the charge conjugate spinor $\psi_c$ is given by $$\psi_c \mapsto \bar{\hat{g}}\cdot\psi_c, \ \pmatrix{\bar{T} & \vec{V} \cr 0 & 1 \cr}\cdot \pmatrix{-\bar{v} \cr \bar{u} \cr}\pmatrix{\vec{x} \cr t}=e^{{{im}\over{\hbar}}(\vec{V}\cdot\vec{x}+{{1}\over{2}}\vert \vec{V} \vert ^2t)}\bar{T}\pmatrix{-\bar{v}(\pi(\bar{T})\vec{x}+\vec{V}t,t) \cr \bar{u}(\pi(\bar{T})\vec{x}+\vec{V}t,t)}. \eqno{(15)}$$ Finally, the galilean transformations of the electromagnetic potential $(\phi, \vec{A})$ and the magnetic field $\vec{B}$ are $$\phi(\vec{x},t)=\phi^\prime(\vec{x}^\prime, t^\prime), \ \vec{A}(\vec{x},t)=R\vec{A}^\prime(\vec{x}^\prime,t^\prime), \ \vec{B}(\vec{x},t)=R\vec{B}^\prime(\vec{x}^\prime,t^\prime) \eqno{(16)}$$ with $\vec{x}=R\vec{x}^\prime+\vec{V}t^\prime$ and $t=t^\prime$. 

\

{\it Remark}: Representations associated with different values of the mass are inequivalent. $^6$

\

3. {\bf Lagrangian formulation and galilean and gauge invariances}

\

The Pauli equations (1) and (5) can be formulated within the lagrangian framework. The lagrangian for equation (1) is $${\cal L}={{i\hbar}\over {2}}((({{\partial}\over{\partial t}}-{{iq}\over{\hbar}}\phi)\psi^\dagger) \psi-\psi^\dagger ({{\partial}\over{\partial t}}+{{iq}\over{\hbar}}\phi)\psi)+{{\hbar ^2}\over{2m}}(\nabla+{{iq}\over{\hbar c}}\vec{A})\psi^\dagger\cdot(\nabla-{{iq}\over{\hbar c}}\vec{A})\psi-{{q\hbar}\over{2mc}}\psi^\dagger\vec{\sigma}\cdot\vec{B}\psi$$ $$={{i\hbar}\over{2}}(\dot{\psi}^\dagger\psi-\psi^\dagger\dot{\psi})+{{\hbar ^2}\over{2m}}\nabla\psi^\dagger\cdot\nabla\psi+{{q^2}\over{2mc^2}}\psi^\dagger\vert \vec{A}\vert ^2\psi+{{i\hbar q}\over{2mc}}(\psi^\dagger\vec{A}\cdot\nabla\psi-\nabla\psi^\dagger\cdot\vec{A}\psi)-{{q\hbar}\over{2mc}}\psi^\dagger\vec{\sigma}\cdot\vec{B}\psi+q\psi^\dagger\phi\psi, \eqno{(17)}$$ and equation (1) amounts to the variational equation $${{\delta}\over{\delta\psi^\dagger(\vec{x},t)}}S=0 \eqno{(18)}$$ where $S$ is the action $$S=\int dt\int d^3\vec{x}{\cal L}(\vec{x},t). \eqno{(19)}$$ Under the charge conjugation operation $${\cal L}\to {\cal L}_c=K{\cal L}=-{{i\hbar}\over {2}}((({{\partial}\over{\partial t}}+{{iq}\over{\hbar}}\phi)\psi^\dagger_c)\psi_c-\psi^\dagger_c ({{\partial}\over{\partial t}}-{{iq}\over{\hbar}}\phi)\psi_c)+{{\hbar ^2}\over{2m}}(\nabla-{{iq}\over{\hbar c}}\vec{A})\psi^\dagger_c\cdot(\nabla+{{iq}\over{\hbar c}}\vec{A})\psi_c+{{q\hbar}\over{2mc}}\psi^\dagger_c\vec{\sigma}\cdot\vec{B}\psi_c$$ 
$$=-{{i\hbar}\over{2}}(\dot{\psi}^\dagger_c\psi_c-\psi_c^\dagger\dot{\psi}_c)+{{\hbar ^2}\over{2m}}\nabla\psi^\dagger_c\cdot\nabla\psi_c+{{q^2}\over{2mc^2}}\psi^\dagger_c\vert \vec{A}\vert ^2\psi_c-{{i\hbar q}\over{2mc}}(\psi_c^\dagger\vec{A}\cdot\nabla\psi_c-\nabla\psi^\dagger_c\cdot\vec{A}\psi_c)+{{q\hbar}\over{2mc}}\psi^\dagger_c\vec{\sigma}\cdot\vec{B}\psi_c+q\psi^\dagger_c\phi\psi_c. \eqno{(20)}$$ 

To pass from (17) to (20), the identity $M^\dagger M=1$ is inserted at each term of (17), and the fact that $M\vec{\sigma}M^\dagger=M(\sigma_1,\sigma_2,\sigma_3)M^\dagger=(-\sigma_1,\sigma_2,-\sigma_3)$ is used; then the complex conjugation operation $K$ completes the transformation. 

\

The total action for the particle-antiparticle system is $$S_{tot}=S+S_c=\int dt \int d^3 \vec{x}({\cal L}(\vec{x},t)+{\cal L}_c(\vec{x},t)) \eqno{(21)}$$ and equation (5) is obtained from $S_{tot}$ or $S_c$ as $${{\delta}\over {\delta \psi^\dagger_c(\vec{x},t)}}S_{tot}={\delta \over {\delta \psi^\dagger_c(\vec{x},t)}}S_c=0. \eqno{(22)}$$

\

The lagrangian ${\cal L}$, and therefore the equation (1), are invariant under the galilean transformations (13), (15) and (16) for $\psi$, $\psi_c$, and $(\phi, \vec{A})$ and $\vec{B}$, respectively. To prove it, we use the facts that $\nabla=R^{-1}\nabla^\prime$ where $\nabla={{\partial}\over{\partial \vec{x}}}$ and $\nabla^\prime={{\partial}\over {\partial\vec{x}^\prime}}$, and ${{\partial} \over{\partial t}}={{\partial} \over{\partial t^\prime}}-R^{-1}\vec{V}\cdot\nabla^\prime$. If ${\cal L}^\prime$ and ${\cal L}^\prime_c$ are the transformed lagrangian densities for particles and antiparticles, then, from equation (20), $${\cal L}_c(\vec{x},t)=K{\cal L}(\vec{x},t)=K{\cal L}^\prime(\vec{x}^\prime,t^\prime)={\cal L}_c^\prime(\vec{x}^\prime,t^\prime) \eqno{(23)}$$ and therefore the galilean invariance of equation (5) is also proved. 

\

Finally, both ${\cal L}$ and ${\cal L}_c$, and therefore the equations (1) and (5), are gauge invariant under the transformations $\psi\to e^{i\Lambda}\psi$, $\psi_c\to e^{-i\Lambda}\psi_c$, $\phi \to \phi-{{\hbar}\over{q}}{{\partial}\over{\partial t}}\Lambda$ and $\vec{A}\to \vec{A}+{{\hbar c}\over{q}}\nabla\Lambda$, where $\Lambda$ is an arbitrary differentiable function of $(\vec{x},t)$.

\

{\bf Acknowledgement}

\

The author was partially supported by the project PAPIIT IN103505, DGAPA-UNAM, M\'exico. 

\

{\bf References}

\

1. A. Cabo, D. B. Cervantes, H. P\'erez Rojas and M. Socolovsky, Remark on charge conjugation in the nonrelativistic limit, {\it International Journal of Theoretical Physics} (to appear); arXiv: hep-th/0504223.

\

2. J. D. Bjorken, and S. D. Drell, {\it Relativistic Quantum Mechanics}, Mc Graw-Hill, New York (1964): p. 11.

\

3. M. Socolovsky, The CPT Group of the Dirac Field, {\it International Journal of Theoretical Physics}  {\bf 43}, 1941-1967 (2004); arXiv: math-ph/0404038.

\

4. V. B. Berestetskii, E. M. Lifshitz, and L. P. Pitaevskii, {\it Quantum Electrodynamics, Landau and Lifshitz Course of Theoretical Physics, Vol. 4}, 2nd. edition, Pergamon Press, Oxford (1982): p. 45.

\

5. J. A. de Azc\'arraga and J. M. Izquierdo, {\it Lie groups, Lie algebras, cohomology and some applications in physics}, Cambridge University Press, Cambridge (1995): p. 155.

\

6. S. Sternberg, {\it Group Theory and Physics}, Cambridge University Press, Cambridge (1994): p. 49.

\

\

\

\

\

e-mail: socolovs@nucleares.unam.mx, somi@uv.es

\

\end